\documentclass[
twocolumn,aps,prl,showpacs,superscriptaddress
]{revtex4-1}

\usepackage{graphicx,color,hyperref}

\usepackage{amsmath}
\usepackage{amsfonts}
\usepackage{amssymb}

\begin{document}

\title{
Quantum criticality of granular SYK matter
}

 \author{Alexander Altland}

\affiliation{Institut f\"ur Theoretische Physik, Universit\"at zu K\"oln,
Z\"ulpicher Stra\ss e 77, 50937 K\"oln, Germany}

\author{Dmitry Bagrets}

\affiliation{Institut f\"ur Theoretische Physik, Universit\"at zu K\"oln,
Z\"ulpicher Stra\ss e 77, 50937 K\"oln, Germany}

\author{Alex Kamenev}

\affiliation{W. I. Fine Theoretical Physics Institute and School of Physics and Astronomy, University
of Minnesota, Minneapolis, MN 55455, USA}

\date{\today}


\begin{abstract}
We consider granular quantum matter defined by Sachdev-Ye-Kitaev (SYK) dots coupled via random one-body hopping. Within the framework of Schwarzian field theory, we identify a zero temperature quantum phase transition between an insulating phase at weak and a metallic phase at strong hopping. The critical hopping strength scales inversely with the number of degrees of freedom on the dots. The increase of temperature out of either phase induces   a crossover into a regime of strange metallic behavior. 
\end{abstract}

\maketitle

\noindent \emph{Introduction:} Despite decades of  research, our understanding of  strongly correlated (`non-Fermi
liquid') quantum matter with metallic parent states remains incomplete. A universal
feature of these materials is that seemingly incongruent phases of matter ---
superconducting, insulating, poorly conducting, metallic, etc. ---  coexist in close
parametric proximity to each other\cite{Fradkin2015}. The understanding of this
diversity of competing phases, which finds its most prominent manifestations in the
physics of the cuprates\cite{Keimer2015} or heavy fermion materials\cite{Si2010},
requires universal blueprints of correlated fermion matter transcending the Landau
quasiparticle paradigm. Recently, systems of coupled Sachdev-Ye-Kitaev
(SYK)\cite{kitaev2015talk,Sachdev-Ye,song2017strongly,Gu2017,chen2017competition,berkooz2017higher,Shenoy2018,Cai2018,Khveshchenko2018,Xu2017,Jian2017,Kim2019,Xu2019}
quantum dots have gained popularity in this context. What makes these systems
interesting is that a hallmark of many correlated fermion materials --- crossover
from a strange metal (SM) phase to a Fermi liquid (FL) upon lowering temperatures ---
is generated within a very simple mean field picture\cite{song2017strongly}, which
assumes the individual SYK cells to contain a thermodynamically large number $N\to
\infty$ of quantum particles. In this paper, we do not take this limit and explore
what happens in `mesoscopic' systems where $N$ is large but finite. Our main finding
is that  the phase diagram becomes significantly more interesting and now features a
zero temperature insulator--FL transition at a critical value of the inter-dot
coupling inversely proportional to $N$. Extending the analysis to finite
temperatures, we find an insulator/SM/FL phase separation as shown in (see Fig.1).
Competitions of this type are seen in many contexts, indicating that the mesoscopic
SYK network may capture essential ingredients  for the phenomenological description
of the correlated fermion matter.

The SYK model\cite{kitaev2015talk,Sachdev-Ye} is a system of $N$ Majorana fermions,
$\eta_i$, $i=1\dots,N$, subject to an all--to--all four fermion interaction
$H_{\mathrm{SYK}}=(1/4!)\sum_{ijkl}^N J_{ijkl}\eta_i\eta_j\eta_k\eta_l$ with Gaussian
distributed matrix elements $J_{ijkl}$ of variance $3! J^2/N^3$. The system can be seen
as a spatially local, zero dimensional paradigm of strongly interacting quantum
matter: In the limit $N\to \infty$, the absence of a single particle term in the
Hamiltonian implies that the fermion operators  carry dimension
$[\text{time}]^{-1/4}$, in marked distinction to the FL dimension $-1/2$. This
motivates the extension to a $d$-dimensional array of nearest neighbor coupled non
Fermi liquid cells. In view of the inherent randomness, it is natural to model the
coupling by one-body operators $H_\mathrm{T} = (i /2)\sum_{\langle
ab\rangle,ij}V^{ab}_{ij}\eta_i^a \eta_j^b$, where $a,b$  label the individual dots,
and $V^{ab}_{ij}$ are Gaussian distributed with  variance $ V^2/N$. Importantly, this
coupling is a {\em relevant} perturbation of dimension $[\int d\tau \eta
\eta]=1-2\times1/4 =+1/2$. It implies a crossover from a non-FL `strange' metal 
at high temperatures to a conventional, yet strongly renormalized, FL metal at low
temperatures\cite{song2017strongly}. 

\begin{figure}[t]
\vskip -.5cm
\hskip -1cm \includegraphics[width=8cm]{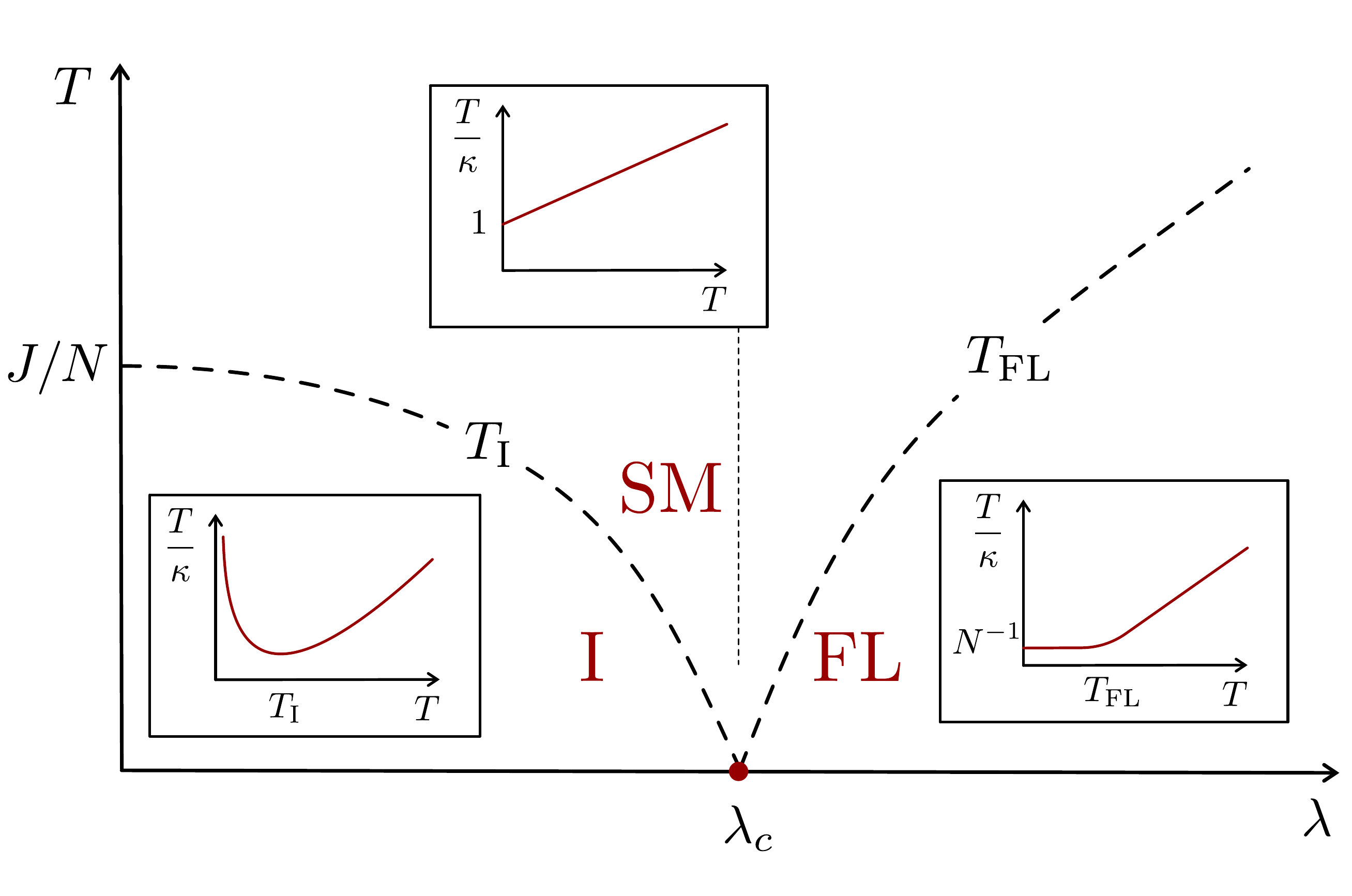}
\vskip -.5cm
\caption{Phase diagram of SYK array: $T$ vs. dimensionless hopping strength $\lambda=(NV/J)^2$. At a critical value, $\lambda_c=8/Z$,  the system undergoes a zero temperature metal--insulator QPT. The two lines $T_\mathrm{I}(\lambda)$ and  $T_\mathrm{FL}(\lambda)$ mark insulator (I) to strange metal (SM) and FL to SM crossovers, correspondingly. The insets shows thermal resistivity $T/\kappa(T)$ vs. $T$ for $\lambda<\lambda_c$, $\lambda=\lambda_c$ and $\lambda>N>\lambda_c$.  } 
\label{fig1}
\end{figure}

The above scenario makes reference to the engineering dimensions of the fermion
operators and becomes valid in the thermodynamic limit. However, for finite $N$, very different behavior at
low temperatures is expected. The non-FL nature of an isolated SYK dot manifests
itself in an infinite dimensional `conformal'
symmetry\cite{kitaev2015talk,Maldacena16,CommentsSYK16,Bagrets-Altland-Kamenev2016,bagrets2017power,Kitaev}
under continuous reparameterizations of time. The above scaling dimension $-1/4$
reflects the breaking of this symmetry at the large $N$ mean-field level. However, as
temperature is lowered below the energy scale $J/N$, strong Goldstone fluctuations
associated to the conformal symmetry ensue, and effectively change the dimension of
the fermion operator to
$-3/4$\cite{Bagrets-Altland-Kamenev2016,bagrets2017power,Mertens2017,Mertens2018}. In
this low energy regime, a single particle perturbation has dimension $1-2\times
3/4=-1/2$ and now is RG \emph{irrelevant}. 

This dimensional crossover implies  a competition between inter-dot couplings and
intra-dot quantum fluctuations: depending on the bare strength of the coupling,
Goldstone modes are either suppressed, or render the inter-dot coupling irrelevant.
This implies the existence of a  metal-insulator quantum phase transition (QPT)
separating a phase of a strongly coupled FL from an insulating phase of essentially
isolated dots. Below, we will explore this QPT within the framework of an effective
low energy field theory describing granular SYK matter in terms of two coupling
constants, representing intra-dot interaction and inter-dot coupling strength,
respectively. We will demonstrate the renormalizability of the theory and from the
flow of coupling constants (cf. Fig.~\ref{fig2} below) derive the manifestations of
quantum criticality in two temperature scales marking an insulator/SM and FL/SM
crossover at weak and strong coupling, respectively, cf. Fig.~\ref{fig1}.

Before turning to the discussion of the model we note that reference~\cite{Feigelman2018} applied similar reasoning to predict
a non-FL/FL phase transition for an isolated SYK dot subject to a one-body perturbation.
We will comment on this result in relation to the I/FL  transition in the array
geometry after developing the proper theoretical framework. On general
grounds we also expect similar physics in models of interacting {\em complex}
fermions, the SY model\cite{kitaev2015talk,Sachdev-Ye,song2017strongly}. However, the
presence of $\mathrm{U}(1)$-mode associated with particle number conservation in the
SY system makes the theory more complicated.  We here  prefer to sidestep this
complication and expose the relevant physics within the SYK framework, unmasked by
the $\mathrm{U}(1)$ phase fluctuations~\cite{altland2006}. In this system of
electrically neutral Majorana fermions, thermal conductivity, $\kappa(T)$, is the
main signature of transport, and  from the Wiedemann-Franz law we infer that the
ratio $T/\kappa$ plays a role analogous to the electrical resistivity of complex
fermion matter. We find that  in the insulating phase it exhibits a minimum before
diverging  at small $T$ as $T/\kappa(T)\propto 1/T$ (cf. bottom left inset in
Fig.~\ref{fig1}). In the SM (FL) phase $T/\kappa$  ratio exhibits   $T$-linear
(approximately $T$-independent) behavior, respectively.

{\em The model:} we consider a system described by the Hamiltonian\cite{song2017strongly}
\begin{equation}
				\label{eq:Hamiltonian}
H=\frac{1}{4!}\sum\limits_a\sum\limits_{ijkl}^N J^a_{ijkl}\eta^a_i\eta^a_j\eta^a_k\eta^a_l  +
\frac{i}{2} \sum\limits_{\langle ab\rangle}\sum\limits_{ij}^N  V^{ab}_{ij}\eta_i^a \eta_j^b,
\end{equation}
where the mutually  uncorrelated Gaussian distributed coefficients  $J^a_{ijkl}$ and
$V^{ab}_{ij}$ have been specified above. Following a standard
procedure\cite{Maldacena16,CommentsSYK16,Bagrets-Altland-Kamenev2016,bagrets2017power,Kitaev},
the theory averaged over the coupling constant distributions is described by an
imaginary time functional $Z= \int\! D(G,\Sigma)\exp(-S[G,\Sigma])$, where
$G=\{G^a_{\tau_1,\tau_2}\}$ and $\Sigma=\{\Sigma^a_{\tau_1,\tau_2}\}$ are time
bi-local integration fields playing the role of the on-site SYK Green function and
self-energy, respectively. The action $S[G,\Sigma]\equiv \sum_a
S_0[G^a,\Sigma^a]+\sum_{\langle ab\rangle}S_{\mathrm{T}}[G^a,G^b]$, contains the
`$G\Sigma$-action', $S_0$, of the individual dots, and a tunneling action
$S_{\mathrm{T}}[G^a,G^b]=\frac 12 NV^2\int\!\!\int\! d\tau_1 d\tau_2 \,
G^a_{\tau_1,\tau_2}G^b_{\tau_2,\tau_1}$  describing the nearest neighbor hopping.
Here, we omit  a replica structure\cite{Wang2018} technically required to perform the
averaging, but inessential in the present context.

While the explicit form of the $G\Sigma$-action\footnote{The $G\Sigma$-action
describing an SYK dot after averaging over disorder reads\cite{Maldacena16,CommentsSYK16,Bagrets-Altland-Kamenev2016,bagrets2017power,Kitaev}  $S_0[G,\Sigma]=-{N\over
2} \mathrm{tr}\ln\left(\partial_\tau +
\Sigma\right)-\frac{N}{2}\int d\tau_1
d\tau_2(G_{\tau_1,\tau_2}\Sigma_{\tau_2,\tau_1}+\frac{J^2}{4}(G_{\tau_1,\tau_2})^4)$.}
will not be needed, the following points are essential: (i) the action $S_0$ possesses an exact $\mathrm{SL}(2,R)$-invariance (see below) and approximate invariance under reparameterizations of  time\cite{kitaev2015talk,Maldacena16,CommentsSYK16,Bagrets-Altland-Kamenev2016,bagrets2017power,Kitaev},  $h:S^1\to S^1, \tau \mapsto h(\tau)$, where $h$ is a diffeomorphism of the circle, $S^1$, defined by imaginary time with periodic boundary conditions onto itself.
The infinite dimensional symmetry group $\mathrm{diff}(S^1)$ of these transformations
is generated by a Virasoro algebra, hence the denotation `conformal'. (ii) The
symmetry is subject to a weak explicit breaking by the time derivatives present in
the action $S_0$. For low energies, the corresponding action cost  is given by\cite{kitaev2015talk,Maldacena16,CommentsSYK16,Bagrets-Altland-Kamenev2016,bagrets2017power,Kitaev,stanford2017fermionic,Blommaert2018,Lam2018}
$S_0[h]= -m \int_0^\beta d\tau \{h,\tau\}$, where  
$\{h,\tau\}\equiv
\big(\frac{h^{\prime\prime}}{h^\prime}\big)^\prime-\frac{1}{2}\big(\frac{h^{\prime\prime}}{h^\prime}\big)^2$ is the  Schwarzian derivative, and the proportionality
 $m\propto N/J$ of the coupling constant indicates that quantum reparameterization fluctuations become stronger for small $N$.  For temperature scales $T< m^{-1}$ even large deviations, $h$, away from
$h(\tau)=\tau$ may have low action. This marks the entry into a low temperature
regime dominated by strong reparameterization fluctuations. Finally, (iv) the
mean-field Green function $G_{\tau_1,\tau_2}= |\tau_1-\tau_2|^{-1/2}$ (the square
root dependence  reflects the non-FL dimension of the fermions) transforms under
reparameterizations as
\begin{equation}
						\label{eq:GF}
G_{\tau_1,\tau_2}\to G_{\tau_1,\tau_2}[h] =  \left( \frac{h'_1 h'_2}{[h_1 -  h_2]^{2}}\right)^{1/4}, 
\end{equation}
where $h_i\equiv h(\tau_i)$ and $h'_i\equiv dh(\tau)/d\tau|_{\tau=\tau_i}$.
For an isolated dot, integration over the $h$-fluctuations effectively changes the Green function to $\langle G_{\tau_1,\tau_2}[h]\rangle_h \stackrel{mT\ll 1}\longrightarrow m |\tau_1-\tau_2|^{-3/2}$, corresponding to a  change of the fermion operator dimension to $-3/4$\cite{Bagrets-Altland-Kamenev2016,bagrets2017power,Mertens2017,Mertens2018,Feigelman2018}.

The effective low-energy \emph{lattice Schwarzian theory}  is formulated in terms of
the reparameterizations $h^a(\tau)$  on different dots. Its action  $S[h] = S_0[h] + S_{\mathrm{T}}[h]$, is defined through
\begin{align}
						\label{eq:lattice-ssh}
S_0[h]&= - m\sum_a \int d\tau \, \{h^a,\tau\},\\
S_{\mathrm{T}}[h]&=- w \sum_{\langle ab\rangle}\! \int\!\!\!\!\int\!  d\tau_1 d\tau_2 \left(\!  \frac{{h'^a_1} {h'^a_2}}{[h_1^a - h_2^a]^{2}} \times\frac{{h'^b_1} {h'^b_2}}{[h_1^b - h_2^b]^{2}}\! \right)^{\!\! 1/4} \!\!,
\nonumber
\end{align}
where  $m$ and $w$ are parameters with dimensions of [time] and [energy], and  bare values $m\propto N/J$ and $w\propto NV^2/J$. 

A hallmark of the  lattice Schwarzian action, $S[h]$,  is its invariance under actions of ${\rm SL}(2,R)$,
where the group is represented via the  M\"obius transformations
$h(\tau)=\frac{\alpha
\tau+\beta}{\gamma\tau +\delta}$ with $\alpha\delta-\beta\gamma=1$. 
This shows that the $h$-transformations to be integrated cover the
coset space $\mathrm{diff}(S^1)/{\rm SL}(2,R)$. The action itself is
built  from the  two simplest ${\rm SL}(2,R)$ invariant blocks: local $\{h,\tau\}$ and
bi-local ${h'_1} {h'_2}/[h_1 - h_2]^{2}$.  Maintained ${\rm SL}(2,R)$ symmetry 
imposes a stringent condition on the behavior of the theory under renormalization. A
successive integration over $h$-transformations must leave the local
and bi-local terms form invariant (multi-point terms may be generated but
are irrelevant). The invariance condition thus implies that the  renormalization results in a flow of the two couplings $m$ and $w$. 

\emph{RG analysis:}  we decompose  fluctuations into 'fast' and 'slow' as  $h(\tau) =
f(s(\tau)) \equiv (f \circ s)(\tau)$, where
 $f$ and $s$ are fluctuations in the frequency  range
$[\Lambda, J]$ and $[0,\Lambda]$, and $\Lambda$ is a running cutoff energy\footnote{In passing we note that the separation
of fast and slow fluctuations for \emph{diffeomorphic}  maps is not straightforward.
For example, representations via superpositions of Fourier modes generally violate
the injectivity required of a reparameterization. However, as with many other RG
procedures, our recursive procedure below relies only on few principal differences
between slow and fast fluctuations and does not require a formal separation.}. We then integrate out the fast modes $f(s)$,
and rescale time $\tau\to \tau J/\Lambda$ to restore the UV cutoff $\Lambda\to J$.  
Consider first the case $m^{-1}< \Lambda<J$, where the reparameterization fluctuations are suppressed. 
The RG flow is then governed by the `engineering' dimensions, resulting in:
\begin{equation}
						\label{eq:RG-bare}
\frac{d\ln m}{dl}=-1;\quad\quad \frac{d\ln w}{dl} = +1,
\end{equation}
where $l=\ln(J/\Lambda)$. For $T>J/N$ this flow should be terminated when either
$\Lambda$ reaches $T$, or $V(l)\sim \sqrt{w(l)}$ reaches the UV cutoff $J$. This
defines the temperature scale $T_\mathrm{FL}=V^2/J$,  separating the high temperature
SM and low temperature FL.  In the SM phase $w(T)=NV^2/T$ and  $T/\kappa(T)\propto J/w(T)
\propto T/(N T_\mathrm{FL})$ \cite{song2017strongly}, while in FL the thermal
resistivity saturates at $ T/\kappa(T)\propto 1/N$.

We now turn to the regime of strong reparameterization fluctuations. When 
$\Lambda$ reaches $J/N$, $m(l)=m(0)e^{-l}$ reaches the inverse UV cutoff $m(l)\approx
1/J$. To proceed with the further renormalization, we employ the Schwarzian chain
rule
\begin{equation}
					\label{eq:chain}
\{f\circ s,\tau\}=(s')^2\{f,s\}+\{s,\tau\}, 
\end{equation}
to obtain the action: $S_0[f \circ s] = S_0^{\rm fast}[f,s] + S_0[s]$,
where 
the 'fast' Schwarzian action has a time-dependent mass
$m(s)\equiv  m s^{a\prime }$.
At  lowest order in $w$ one needs to average the coupling action
$S_{\mathrm{T}}[f\circ s]$ over the fast fluctuations.
A straightforward application of the chain rule to  the Green functions, Eq.~(\ref{eq:GF}), shows that  
\begin{align}
\label{eq:GF-chain} 
G_{\tau_1,\tau_2}[f\circ s]=G_{s_1,s_2}[f](s'_1s'_2)^{1/4}, 
\end{align}
so that $\langle S_{\mathrm{T}}[f\circ s]\rangle_f
\propto  \langle G_{s_1,s_2}[f^a]\rangle_{f^a} \times \langle G_{s_2,s_1}[f^b]\rangle_{f^b}
$ splits into two fast averages. These expressions can be evaluated with the help of exact results \cite{Bagrets-Altland-Kamenev2016,Mertens2017} for the 2-point propagator of the Schwarzian theory.
Referring to 
the supplementary material  for details~\footnote{
See  Supplementary Material for the technical background on the lattice Schwarizian field theory.
}, we note the asymptotic expressions ($s_{12}\equiv s_1-s_2$): 
\begin{equation}
						\label{eq:GF-asymptotic}
\langle G_{s_1,s_2}[f]\rangle_f  \simeq \begin{cases}
|s_{12}|^{-1/2}, &  \!\! s_{12}< m;\\   
\!\sqrt{m(s_1)m(s_2)}  |s_{12}|^{-3/2}, &\!\!  m\! <\! s_{12}\!<\! \Lambda^{-1}; \\
m\Lambda |s_{12}|^{-1/2}, &\!\!  \Lambda^{-1} < s_{12}.    
\end{cases} 
\end{equation}
This equation implies that the double time integral  in the averaged
tunneling action $\langle S_{\mathrm{T}}[f\circ s]\rangle_f\equiv S_{\mathrm{int}}+S_{\mathrm{long}}$ gets different contributions from intermediate ($m <
\tau_{12}< \Lambda^{-1}$) and long time differences ($\tau_{12}>\Lambda^{-1}$). In processing the former, we  use the general Taylor expansion ($\tau = (\tau_1 + \tau_2)/2$)
\begin{align}
                            \label{eq:Schw_expand}
\left( \frac{s'_1 s'_2}{[s_1-s_2]^2}  \right)^\Delta \approx
\frac{1}{[\tau_1 - \tau_2]^{2\Delta}} + \frac{\Delta}{6} 
\frac{\{s(\tau),\tau\} }{[\tau_1 - \tau_2]^{2\Delta-2}}  + \dots,  
\end{align}
with $\Delta=3/4$ to process the rational functions of the slow fields appearing upon
substitution of Eqs.~\eqref{eq:chain} and ~\eqref{eq:GF-asymptotic} into the action.
Here, the second term indicates how the non-linear action of the tunneling term
manages to feed back into the Schwarzian action under renormalization. Carrying out
the details of the RG step (see supplementary material) and rescaling time to retain the value of the cutoff, $\Lambda$, we find that the integration
over the intermediate time domain changes the coefficient of the local action as
$m\to m(l)\equiv e^{-l}(m+
\frac{Z}{4} w m^2 l)$. The complementary integration over large time differences conserves the form of the tunneling action but changes the coupling constant as $w\to w(l) =e^l w(m\Lambda)^2 = e^lwe^{-2l}$


From these results, RG equations  are obtained by differentiation over $l$ and putting $l=0$. This leads to 
\begin{equation}
 						\label{eq:RG-1order}
\frac{d\ln m}{dl} =-1 + \frac{Z}{4} w m;\quad\quad \frac{d\ln w}{dl} = +1-2.
\end{equation}
The second equation reflects the aforementioned change of the dimension of $w$ from
$+1$ to $-1$. While Eqs.~(\ref{eq:RG-bare}) are applicable for $mJ\gg 1$, the new set
of the RG equations (\ref{eq:RG-1order}) is derived in the opposite limit $mJ\ll 1$.
(Indeed, this is the condition under which the exact expressions for the
propagator\cite{Bagrets-Altland-Kamenev2016,Mertens2017} can be reduced to the
asymptotic expressions (\ref{eq:GF-asymptotic}), see the supplementary material.)

\begin{figure}[t]
\includegraphics[width=8cm]{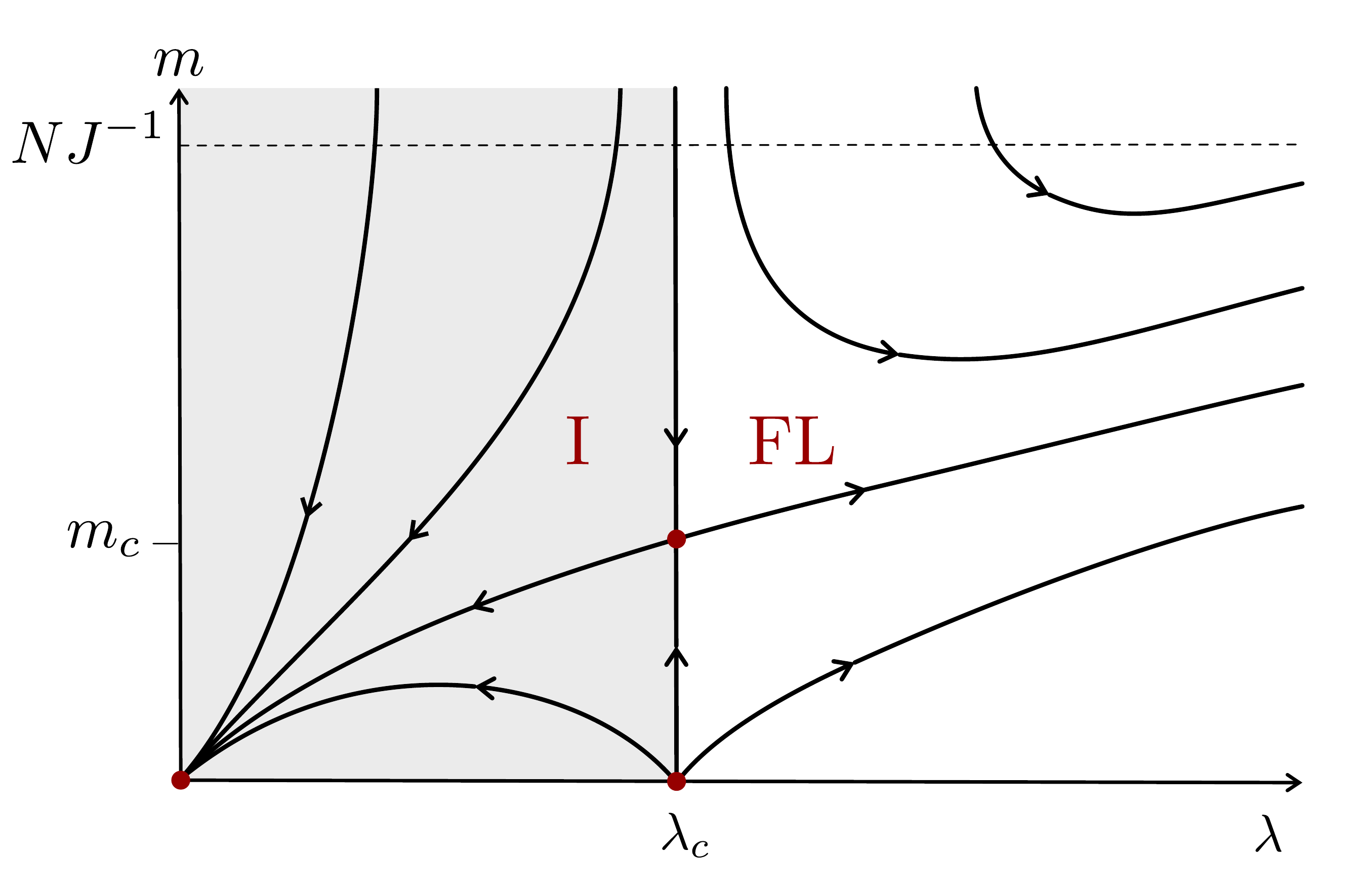}
\caption{RG flow in the plane of couplings $(\lambda=mw,m)$; here $\lambda_c=8/Z$ and $m_c={\cal O}(1)/J$. The initial values are $m(0)=N/J$ and $\lambda(0)=(NV/J)^2$.}
\label{fig2}
\end{figure}

\begin{figure}[b]
\includegraphics[width=5cm]{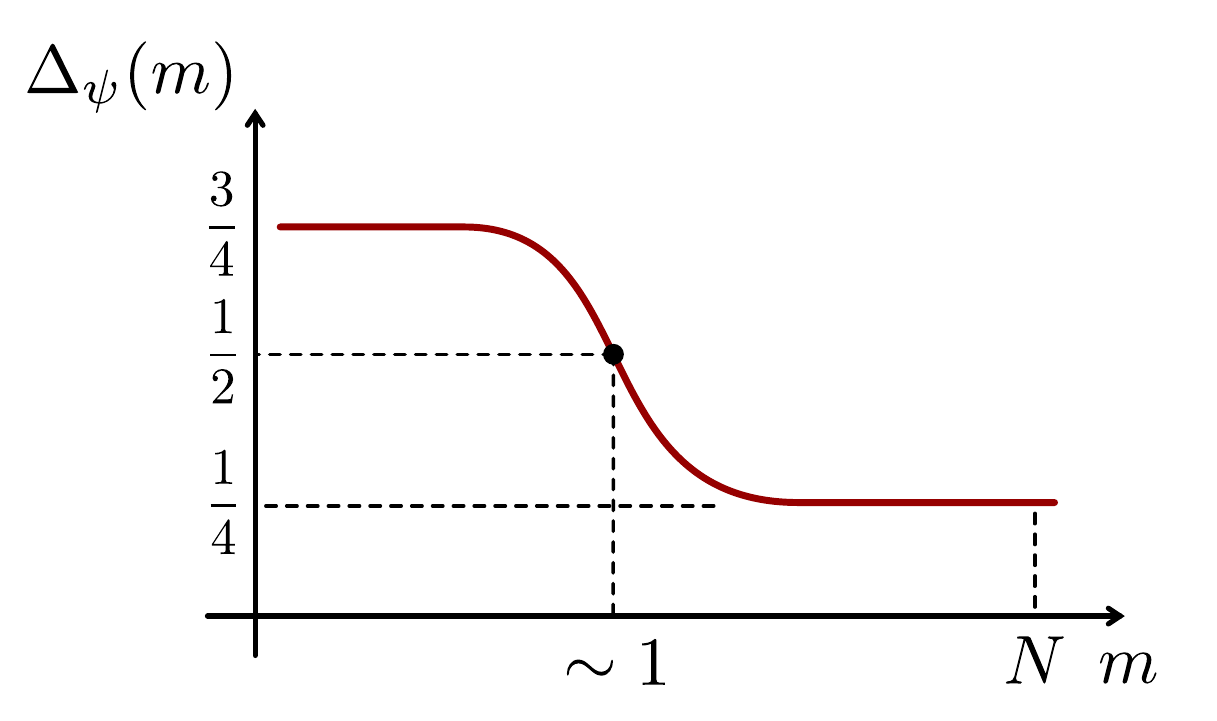}
\caption{
The log-linear plot of the effective scaling dimension of fermion operators $\Delta_\psi$, as a function of the running scale $m$ measured in units of the interaction strength $J$. For its exact definition in terms of the two-point function of the Schwarzian theory, we refer to the Supplemental Material.}
\label{fig3}
\end{figure}

{\em Analysis of the RG:} we first note that the limiting forms of the scaling equations, Eqs.~(\ref{eq:RG-bare}) and (\ref{eq:RG-1order}), admit a closed representation in the dimensionless variable $\lambda\equiv wm$. In the regime $mJ\gg 1$ one has $d\ln \lambda/dl=0$, while for $mJ\ll 1$:  
\begin{align}
                            \label{eq:RG1}
    \frac{d\ln \lambda}{dl}= \left(\frac{Z}{4}\, \lambda  -2\right). 
\end{align}  
This equation exhibits an unstable fixed point  $\lambda_c=\frac{8}{Z}$, marking  a
transition between a FL phase at $\lambda>\lambda_c$ and an insulating one at
$\lambda<\lambda_c$. Since $\lambda(0)\sim (NV/J)^2$, one finds  $V_c\sim
J/\sqrt{Z}N$, inversely proportional to $N$, as stated in the introduction. Notice
that according to Eq.~(\ref{eq:RG-1order}), $d\ln m/dl|_{\lambda=\lambda_c}=+1$,
opposite to Eq.~(\ref{eq:RG-bare}). The only way to reconcile the two limits is to
have another fixed point at $m_c\sim 1/J$. The resulting two parameter RG diagram in
the plane $(\lambda,m)$ is shown in Fig.~\ref{fig2}.    To first order in an
expansion in $w$, but arbitrary $m$,  this diagram may be derived from exact
expressions for $\langle G_{s_1,s_2}[f]\rangle_f$, see supplementary material for
details. In particular, the RG equation for $w$ becomes $d \ln w/dl = 2 - 4 \Delta_\psi(m)$,
where $\Delta_\psi(m)$ is the effective $m$-dependent scaling dimension of the fermion, see Fig.~\ref{fig3}.
The analysis of higher orders in $S_\mathrm{T}$ shows that the actual small
parameter of the perturbative expansion is $Z\lambda $. Therefore, the fixed point is
actually out of the perturbatively controled  regime and may not be used for
quantitative evaluation of critical indices. However, second order
calculations~\footnote{A. Altland, D. Bagrets, and A. Kamenev, in preparation} show
that RG flow keeps its qualitative form, Fig.~\ref{fig2}.

The FL part of the RG diagram, Fig.~\ref{fig2}, is well described by
Eqs.~(\ref{eq:RG-bare}) and the physics of the array is the one discussed in
Ref.~\cite{song2017strongly}. The only addition is that the crossover temperature
$T_\mathrm{FL}(\lambda)\to 0$, when $\lambda\to \lambda_c$, Fig.~\ref{fig1}. This is
due to the fact that for $\lambda\approx \lambda_c$ the flow spends a long ``time''
in the vicinity of the $(\lambda_c,m_c)$ fixed point, thus reaching progressively
lower $T$. In the insulating phase, $\lambda\to 0$ and thus according to $dw/dl=-w$
(Eq.~(\ref{eq:RG-1order})) and $w\sim V^2$, $V(T)\propto T^{1/2}$. 
The diminishing of the inter-dot coupling at low temperatures implies that  second order perturbation
theory in $V(T)$ may be applied to evaluate the thermal conductivity $\kappa(T)$. 
Therefore one finds $\kappa(T)/T \propto |V(T)|^2\propto T$ in the insulating phase.

To conclude, we have seen that the renormalization procedure indeed preserves the
form of the lattice Schwarzian field theory. This stability  follows from the
conformal relations (\ref{eq:chain}) and (\ref{eq:Schw_expand}), but ultimately is
required by the condition of maintained ${\rm SL}(2,R)$ symmetry. Our ability to
deduce the entire RG flow (for $Z\lambda \lesssim 1$) is owed to the knowledge of the
reparameterization averaged Green function $\langle G[f]\rangle_f$ for  any $m$,
which in turn follows from mapping of the local Schwarzian action to Liouville
quantum mechanics~\cite{Bagrets-Altland-Kamenev2016}.  We
finally note that the RG procedure introduced in this Letter may likewise be applied
to  an isolated SYK dot subject to a random one-body perturbation~\cite{Feigelman2018}.
The most important difference is that the action $S_\mathrm{T}$ is now subject to
only one, and not two different reparameterization modes. This leads to a
set of RG equations~\footnote{We obtained the RG equations of the single dot system as $d \ln m/dl= -1+mw/12$
and $d\ln w/dl = 1-1/2$. For $\lambda=mw$ this implies $d\ln \lambda/dl =
\lambda/12-1/2$ and thus the critical value $\lambda_c=6$.}, different from the present ones in that strength of the one-body term,  $w$, remains always relevant.
At the same time, there is a transition in the scaling of $m$, separating a FL phase ($m\gg 1/J$) 
from a phase of strong quantum fluctuations ($m\to 0$), in line with the prediction of
Ref.~\cite{Feigelman2018}. 

\noindent \emph{Summary ---}  In this work we have shown that, regardless of
dimensionality and geometric structure, an array of SYK dots coupled by one-body
hopping exhibits a zero temperature metal-insulator transition. This phenomenon is
rooted in the conformal invariance of the non-FL states supported by the individual
SYK dots. The presence of this symmetry in turn is a direct consequence of an
asymptotically strong dot-local interaction  and may  transcend the specific model
employed here. A mutually suppressive competition between conformal fluctuations on
the dots and the conformal symmetry breaking tunneling operators implies the
presence of a transition between an insulating and a metallic phase, and a crossover
into a strange metal regime at finite temperatures. Read in this way, the main
message of our study is that phenomenology present in many strongly correlated
materials, may follow from a rather basic principle. Although, the underlying
Schwarzian lattice theory will not be able to describe the specific physics of
realistic quantum materials, it will be intriguing to find out if the universality
class of its phase transition can encompass strong correlations phenomena  
beyond those discussed here.

{\it Acknowledgements ---} We are grateful to M. Feigelman and K. Tikhonov for useful discussions.
Work of AA and DB was funded  by the Deutsche Forschungsgemeinschaft (DFG, German Research Foundation) Projektnummer 277101999 TRR~183 (project A03). 
AK was supported by the DOE contract DEFG02-08ER46482.

\bibliography{SYK}

\vskip 0.5cm
\begin{center}
{\bf Quantum criticality of granular SYK  matter: supplementary material
}\end{center}
\vskip .1cm

In this supplementary material we provide some technical background on the lattice Schwarizian field theory discussed in the main text.

\noindent\emph{Effective action and stationary phase solutions} --- Averaging the
Grassmann coherent state path integral representation of the SYK Hamiltonian over the
Gaussian distributions of matrix elements $J^a_{ijkl}$ and $V^{ab}_{ij}$ and
subsequently integrating out the Grassmann variables, one obtains two contributions to the action:
\begin{align} &S_0[G^a,\Sigma^a]=-{N\over 2} \sum_a \Big[
    \mathrm{tr}\ln\left(\partial_\tau + \Sigma^a\right)\nonumber \\ &+ \int d\tau_1
    d\tau_2\left(G^a_{\tau_1,\tau_2}\Sigma^a_{\tau_2,\tau_1}+\frac{J^2}{4}(G^a_{\tau_1,\tau_2})^4\right)\Big];
\end{align} 
and 
\begin{equation}
    S_\mathrm{T}[G^a,G^b] = \frac 12 N\sum_{\langle ab\rangle}V^2 \int d\tau_1 d\tau_2
    G^a_{\tau_1,\tau_2} G^b_{\tau_2,\tau_1},
\end{equation}
where we have suppressed the replica structure of the fields. The global factor $N$ upfront justifies a saddle point approach based on variational solutions for $G$ and $\Sigma$. To zeroth order in $\partial_\tau$ and $V^2$, one finds a family of conformally invariant solutions, parameterized by diffeomorphisms $h(\tau)$: 
\begin{equation}
						\label{eq:suppl:GF}
G_{\tau_1,\tau_2}[h] \!\propto \! \left(\!\frac{h'_1 h'_2}{[h_1 -  h_2]^{2}}\! \right)^{\!1/4}\!\!\!; \quad
\Sigma_{\tau_1,\tau_2}[h] \!\propto \! \left(\!\frac{h'_1 h'_2}{[h_1 -  h_2]^{2}}\! \right)^{\!3/4}\!\!\!. 
\end{equation}  
The terms coupled to $\partial_\tau$ and $V^2$ break the $\mathrm{Diff}(S^1)$ symmetry down
to ${\rm SL}(2,R)$, and  the corresponding action cost is given by Eq.~(3) of
the main text. Eq.~(3) thus defines  the low-energy effective action of the SYK array.

\noindent \emph{RG analysis} --- We now decompose the $h(\tau)$ fluctuations into 'fast' and 'slow' as  $h(\tau)
= f(s(\tau)) \equiv (f \circ s)(\tau)$. The fast part, $f(s)$, includes fluctuations in
the frequency  range $[\Lambda, J]$, and the slow one, $s(\tau)$, the remaining
modes with frequencies less than $\Lambda$. As long as $m^{-1}<\Lambda$, the action cost of fast modes is high and their integration
 has no bearing on the  slow action. Effective renormalization
sets as $\Lambda < m^{-1}$. To first order in $w$ one needs to consider $\langle
S_\mathrm{T}[(f \circ s)]\rangle_f$, which takes the following form (cf. Eq.~(5) of
the main text):
\begin{equation}
								\label{eq:suppl:ST-averag} 
\langle S_\mathrm{T}\rangle_f=  -  w\sum_{\langle ab \rangle}\!\!
 \int\!\!\!\int  
d\tau_1 d\tau_2 \, \langle G_{s^a_1, s^a_2}\rangle_{f^a}  [{s'}^a_1 {s'}^a_2]^{1/4} \times \big( a \to b \big), 
\end{equation}
where the  Green function averaged over the fast degrees of freedom is
\begin{equation}
\label{eq:fast_G}
\langle G_{s_1, s_2}\rangle_{f} =G_f(s_1, s_2) =\Bigl\langle \frac{\left({{f'}(s_1)} {{f'}}(s_2)\right)^{1/4} }
{[f(s_1) - f(s_2)]^{1/2}}\Bigr\rangle_{f},
\end{equation}
and $\langle\ldots\rangle_f$ stands for the integration over the functions $f(s)$
with the weight $S_0^{\rm fast}[f,s] $. Below we will show that this function shows different power law scaling depending on the separation of its arguments (see Fig.~\ref{fig:G_f}):
\begin{figure}[t]
\includegraphics[width=6.5cm]{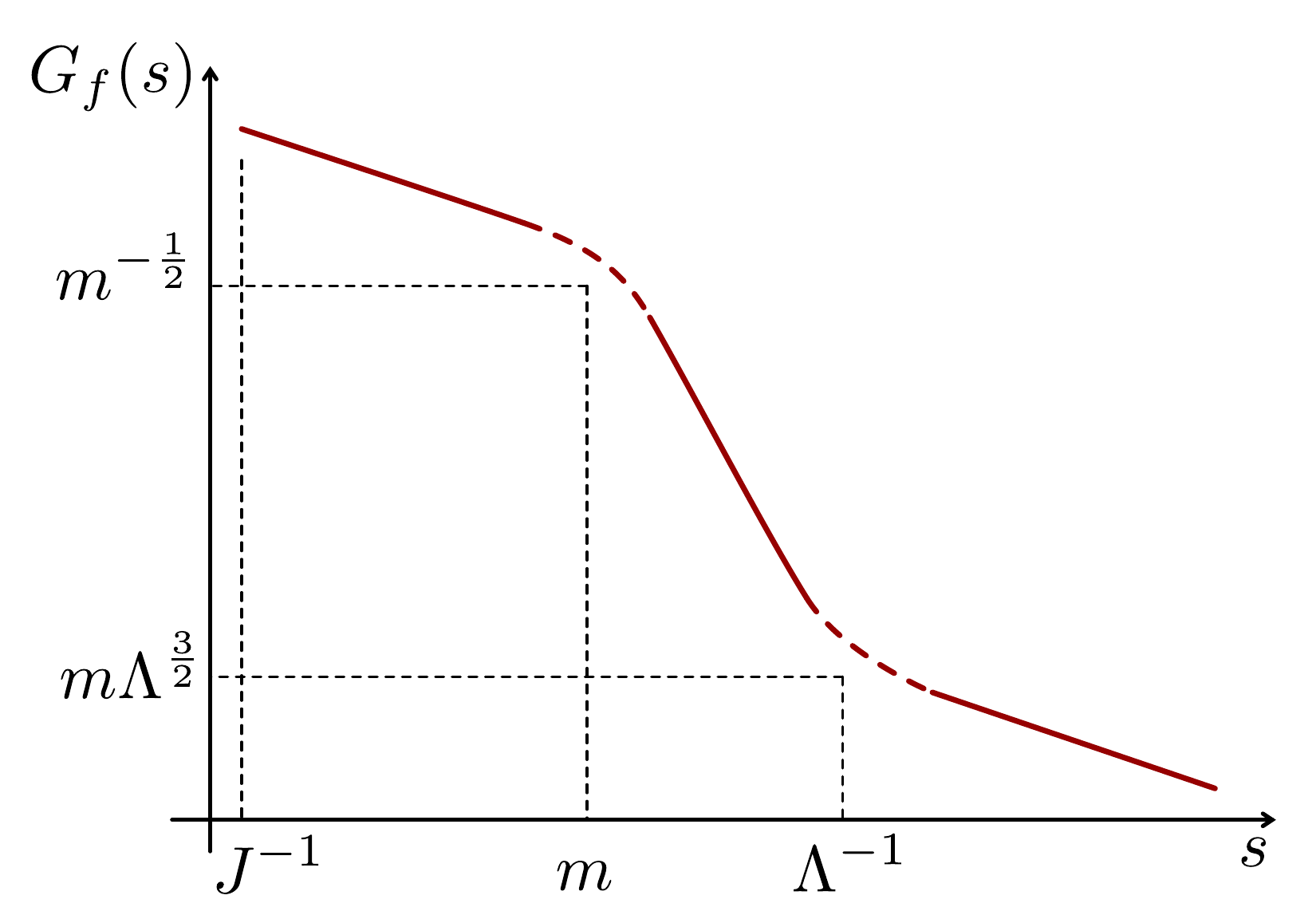}
\caption{The log-log plot of the fast Green's functions $G_{f}(s)$  versus time $s$ used in the RG analysis.}
\label{fig:G_f}
\end{figure}
\begin{equation}
\label{eq:fast_G_asym}
G_f(s_1, s_2) \simeq 7
\frac{{m}^{1/2}(s_1) {m}^{1/2}(s_2)}{ |s_1 - s_2|^{3/2}}, 
\hskip .2cm m < \!|s_1 - s_2| \!< \!\Lambda^{-1}   
\end{equation}
at intermediate time ranges and
\begin{equation}
\label{eq:fast_G_asym_long_t}
G_f(s_1, s_2) =
\frac{m\Lambda}{|s_1 - s_2|^{1/2}},   \qquad |s_1 - s_2| > \Lambda^{-1}
\end{equation}
for long times. We present detailed derivation of these expressions below. The intuition behind them is as follows: the change of the exponent from $-1/2$ to $-3/2$ at  times $>m$ is a result of quantum fluctuations, leading to Liouville quantum mechanics \cite{Bagrets-Altland-Kamenev2016}.  Since  the low-frequency spectrum of fluctuations $f(s)$ is cut off by $\Lambda$, one expects  that the Green function at longer times  ($> \Lambda^{-1} $) 
turns back to its mean-field form with the exponent  $-1/2$. The  suppression
factor $m\Lambda<1$ accounts for the drop of the Greens function in the intermediate
time range. In Eq.~(\ref{eq:fast_G_asym}) we have evenly split the mass
$m(s)=m s'$ between the two times, $s_1$ and $s_2$, which is permissible on account of the assumed slowness of $s$ and leads to manifest ${\rm SL}(2,R)$ invariance of the slow modes action.

As noted in the main text, the average action (\ref{eq:suppl:ST-averag}) acquires  contributions from intermediate and long time differences $\tau_1-\tau_2$. Using the above asymptotic expressions for the Green functions, we obtain, respectively,
\begin{equation}
					\label{eq:suppl:int}
S_\mathrm{int} \!=\! -  w \,m^2\!\! \int\hskip -.95cm \int\limits_{m<|\tau_{12}| <  \Lambda^{-1}} \hskip -.8cm 
d\tau_1 d\tau_2 \!\left(\!  \frac{{s'}^a(\tau_1) {s'}^a(\tau_2)} {|s^a(\tau_1) - s^a(\tau_2)|^{2}}\!\right)^{3/4}\!\!\!\! \times
\Bigl( a \to b\Bigr).
\end{equation}
and \begin{equation}
S_\mathrm{long} \!=\! -  w(m\Lambda\!)^2 \!\!\! \int \hskip -.7cm \int\limits_{|\tau_{12}| >  \Lambda^{-1}} \hskip -.6cm 
 d\tau_1 d\tau_2\!
\left(\!  \frac{{s'}^a(\tau_1) {s'}^a(\tau_2)} {|s^a(\tau_1)\! -\! s^a(\tau_2)|^{2}}\!\right)^{\!1/4}\!\!\!\!\! \times \!
\Bigl( a\! \to\! b\Bigr).
\end{equation}
In the first of these integrals, we use the expansion Eq. (8) of the main text. We
then note that the contribution of lowest order in derivatives comes from the cross
contribution of the first and the second term in the $(a)\times(b)$-product. Turning
to center of mass coordinates, $\tau=(\tau_1+\tau_2)/2$ and
$\tau_{12}=\tau_1-\tau_2$, the integral factorizes into  a contribution of local
Schwarzian form $S_{\mathrm{int}}\to \frac{Z}{4} w m^2 l\sum_a\int_{\Lambda^{-1}}
d\tau \{s^a,\tau\}$, and a logarithmic factor
$l\equiv\ln(1/\Lambda m)= \int_{m}^{\Lambda^{-1}} d\tau_{12}/\tau_{12}
$. Here, $Z$ is the coordination number of the array. We finally rescale the time variable $\tau\to e^l
\tau$, to reset the cutoff $\Lambda^{-1}\to m$, to obtain the original Schwarzian action with a coupling constant $m(l)=e^{-l}(m+
\frac{Z}{4} w m^2 l)$, as stated in the main text. Turning to   the contribution, $S_\mathrm{long}$,  we observe that this one already has the form of the original tunneling action $S_\mathrm{T}[s]$. All that remains to be done is to rescale time which generates the renormalized coupling constant $w(l) =e^l w(m\Lambda)^2 = e^lwe^{-2l}$.

Differentiation of the running constants over $l$ generates the following RG equations:
\begin{equation}
\label{eq:RG_m_small}
\frac{d \ln m}{dl} = -1 + \frac{Z }{4} m w, 
\qquad \frac{d  \ln w}{dl} = -1, \qquad mJ\ll 1.
\end{equation} 
On the other hand, when $mJ\gg 1$ the renormalization is only due to the engineering dimensions,
\begin{equation}
\label{eq:RG_m_high}
\frac{d \ln m}{dl} = -1, 
\qquad \frac{d  \ln w}{dl} = + 1, \qquad mJ\gg 1.
\end{equation} 
A way to interpolate between   the two 
limits~(\ref{eq:RG_m_small}) and (\ref{eq:RG_m_high}) is to  define an effective $m$-dependent scaling dimension of the fermion operators as, 
\begin{equation}
\Delta_\psi(m) = - \frac 12 \frac {d\ln G(s)}{d\ln s}\Biggl|_{s=1/J},
\end{equation} 
where the exact two-point Green's function is~\cite{Bagrets-Altland-Kamenev2016}
\begin{eqnarray}
\label{eq:G2_exact_SM}
&&G(s) \propto \frac{1}{\sqrt m} \int_0^{+\infty} dk\, {\cal M}_2(k) e^{- k^2 s /2m}, \\
&& {\cal M}_2(k) = k \sinh(2\pi k) \Gamma^2(\tfrac 1 4  + i k) \Gamma^2(\tfrac 1 4  - i k). \nonumber
\end{eqnarray}
The function $\Delta_\psi(m)$ smoothly interpolates between $\Delta_\psi = 3/4$ at $mJ \ll 1$ and 
$\Delta_\psi = 1/4$ at $mJ \gg 1$, cf.~Fig.~3 of the main text. 
Using this representation of the two-point function, the RG equations may be derived along the same lines as above, leading to
\begin{eqnarray}
						\label{eq:RG_Delta_psi}
&& \frac{d \ln m}{dl} = -1 + \frac{Z }{4} m w \Bigl( 2 \Delta_\psi(m) - \tfrac 12\Bigr), \\
&& \frac{d  \ln w}{dl} = 1-2\Bigl( 2 \Delta_\psi(m) - \tfrac 12\Bigr).  \nonumber
\end{eqnarray} 
In terms of $m$ and $\lambda=mw$ they read 
\begin{eqnarray}
						\label{eq:RG_Delta_psi-1}
&& \frac{d \ln m}{dl} = -1 + \frac{Z }{4} \lambda \Bigl( 2 \Delta_\psi(m) - \tfrac 12\Bigr), \\
&& \frac{d  \ln \lambda}{dl} = \left(\frac{Z }{4} \lambda -2\right)  \Big( 2 \Delta_\psi(m) - \tfrac 12\Bigr).  \nonumber
\end{eqnarray} 
These equations are valid to the first order in $Z\lambda$, but for arbitrary $mJ$.
They interpolate between the two limits elaborated in the main text. The
corresponding RG flow diagram is presented in the main text as Fig.~2. It contains
the non-trivial hyperbolic fixed point $(\lambda_c,m_c)$, with
$\lambda_c=\frac{8}{Z}$ and  $\Delta_\psi(m_c)=\frac 1 2$. Notice that at the
critical point, the system shows FL scaling.

Linearizing Eqs.~(\ref{eq:RG_Delta_psi-1}) around this fixed point one finds
\begin{equation*}
\frac {d}{dl} \left( \begin{array}{c} \delta \lambda \\ \delta m \end{array} \right) = 
\left( \begin{array}{cc} 1 & 0 \\ m_c/\lambda_c &  \kappa  \end{array} \right)
\left( \begin{array}{c} \delta \lambda \\ \delta m \end{array} \right),
\end{equation*}
where  
\begin{equation*}
\kappa \equiv  4 \, \frac{d \Delta_\psi(m)}{d\ln m}\Biggl|_{m=m_c} = - 0.41.  
\end{equation*}
The two Lyapunov exponents corresponding to the relevant and irrelevant directions  are thus  found as  $\kappa_\mathrm{r} = 1$ and $\kappa_\mathrm{irr} =  \kappa$. The former specifies that the crossover scales $T_\mathrm{I}(\lambda)\propto (\lambda_c-\lambda)$ and   $T_\mathrm{FL}(\lambda)\propto (\lambda-\lambda_c)$ behave linearly near the critical point, see Fig.~1 of the main text. 

\noindent\emph{'Fast' Green function} --- The above analysis relies on
Eqs.~\eqref{eq:fast_G_asym}, \eqref{eq:fast_G_asym_long_t}, or  Eq.~(7) of the main
text. While the general form of these asymptotics follows from qualitative reasoning, the derivation from the Schwarzian theory requires some work. We start from a representation of the Schwarzian action in terms
of the Liouvillian action of a quantum particle  with coordinate $\phi(s) \equiv
\ln f'(s)$~\cite{Bagrets-Altland-Kamenev2016}. This enables one to represent the
'fast' Green function as the following path integral
\begin{equation}
\label{eq:G2_pathI_SM}
G_f(s_1 - s_2) = {\cal Z}^{-1}\! \int\! {\cal D} \phi\, \frac{e^{\tfrac 14 \phi(s_1)} e^{\tfrac 14 \phi(s_2)}}{[\int_{s_1}^{s_2} ds \,e^{\phi(s)} ]^{1/2}}
\,e^{-S_0[\phi]}, 
\end{equation}
with the action 
\begin{equation}
S_0[\phi] =  \frac{m}{2} \int\! ds\, \phi^{\prime 2} + \gamma \int d s \cosh \phi. 
\end{equation}
We have introduced a regulator ($\propto \gamma$), whose role is to
eliminate large negative values of $\phi$, corresponding to slow fluctuations of $f(s)$ (large positive $\phi$'s are cut 
anyways by the Liouville potential, see Eq.~(\ref{eq:sup:H}) below). The value of $\gamma$ will be adjusted below to put the low frequency cutoff at $\Lambda$.  

We employ now a Feynman trick, 
\begin{equation}
\frac{1}{A^{1/2}} = \frac{1}{\sqrt \pi} \int\limits_0^{+\infty} \frac{d\alpha}{\sqrt{\alpha}}\,\, e^{- \alpha A},
\end{equation}
to promote the denominator in Eq.~(\ref{eq:G2_pathI_SM}), $A=\int_{s_1}^{s_2} ds \,e^{\phi(s)} $, to the exponent. This brings the Liouvillian action
 \begin{equation}
 S_\alpha[\phi] = S_0[\phi] + \alpha \int_{s_1}^{s_2}\!d s\,   e^{\phi(s)},
 \end{equation}
which represents a  'quantum quench' problem with the Liouville potential $\alpha e^\phi$, being turned on and off at times $s_1$ and $s_2$. At this point, it is convenient to pass from the path integral formulation to the equivalent  quantum problem governed by the time-dependent Hamiltonian 
\begin{equation}
\label{eq:sup:H}
H_\alpha(s) = - \frac{\partial_\phi^2}{2m}+ \alpha(s) e^\phi + \gamma \cosh \phi, 
\end{equation}
where the parameter $\alpha(s)$ assumes step-wise change  from zero to $\alpha$ at $s=s_1$ and back to zero at $s=s_{2}$. We denote the  eigenstates and energies of the static $\alpha(s)=\alpha$ problem by  $|
\alpha, k^\alpha_n \rangle$ and $E_n^\alpha$, respectively, where $k^\alpha_n$,
$n=1,2,3, \dots$, are effective momenta whose quantization we discuss below. In this language, the Green function~(\ref{eq:G2_pathI_SM}) can be
represented in terms of an exact spectral decomposition,
\begin{equation}
\label{eq:G_f_spectral}
G_f(s_{12}) =\frac{{\cal Z}^{-1}}{\sqrt{\pi}} \int\limits_0^{+\infty} \frac{d\alpha}{\sqrt{\alpha}} 
\sum_{k^\alpha_n}|\langle 0, k^0_0| e^{\phi/4}| \alpha, k^\alpha_n \rangle|^2 
e^{ - E_n^\alpha s_{12}},
\end{equation}
where $s_{12} = s_1 - s_2 >0$ and $|0,k^0_0\rangle$ is the ground state of $H_0$.  

To find an approximate spectrum and eigenstates of $H_\alpha$ we treat it as a quantum well with the width:  \begin{equation}
L_\alpha =  \ln \frac{1}{ m \gamma} + \ln \frac{1}{ m (\alpha + \gamma)},
\end{equation}
inside which the quantum particle is essentially free. The latter gives the energy eigenvalues and momentum quantization (see Ref.~\cite{Bagrets:2016} for a more detailed discussion)
\begin{equation}
E^\alpha_n = \frac{(k^\alpha_n)^2}{2m}, 
\qquad k^\alpha_n = \frac{\pi n }{L_\alpha},\qquad n=1,2,\dots 
\end{equation}
To evaluate the above spectral sum, we need the unit-normalized  eigenstates close to the right boundary of the potential well where they take the form
\begin{equation}
\label{eq:Matrix_El}
\langle \phi | \alpha, k^\alpha_n \rangle = \frac{{\cal N}(k^\alpha_n)}{( 2 L_\alpha)^{1/2}} \, K_{2 i k^\alpha_n}
\left(2 \sqrt{2M(\alpha + \gamma) }e^{\phi/2}\right).
\end{equation}
Here ${\cal N}(k) = {2}/{\Gamma(-2 i k)}$ and $K_{2ik}(z)$ are the generalized Bessel functions. 

We now adjust the strength of the regulator $\gamma$ in a way to match the ground state
energy with the running energy cutoff,  $E_0^0 = \Lambda$. This ensures  that fluctuations at
longer time scales are effectively eliminated. Since $E_0^0 \sim 1/(m
L_0^2)$ and $L_0 \sim \ln \left(1/ m \gamma\right)$, one finds   $\gamma \sim m^{-1} e^{-1/\sqrt{m\Lambda}}$. Within this setting, we next
address  two complementary time regimes: for short and intermediate times, $s \ll
\Lambda^{-1}$,  the relevant energies $E_n^\alpha$ in Eq.~(\ref{eq:G_f_spectral}) are
large as compared to $E_0^\alpha$. Hence, we can use a continuous approximation and
replace the sum over $k_n^\alpha$ by an  integral over a continuous variable $k$.
In this case, the partition sum, ${\cal Z}$, is close to unity. Evaluating the
matrix element $\langle 0, k^0_0| e^{\phi/4}| \alpha, k^\alpha_n \rangle$ and
performing the $\alpha$-integration one arrives~\cite{Bagrets-Altland-Kamenev2016} at  the exact 2-point Green's
function~(\ref{eq:G2_exact_SM}) of the Schwarzian theory, which means that the presence of the cutoff is inessential in this case. In particular, for short
times, $s \ll m$, reparametrizations are not effective and $G_f(s)$ retains its mean
field value $1/s^{1/2}$, while at intermediate ones it crosses over to $m/s^{3/2}$.

In the complementary domain of long times, $s \gg \Lambda^{-1}$, we  employ Eq.~(\ref{eq:G_f_spectral}), where 
for $\beta \gg 1/E_0^0$, one finds 
\begin{equation}
{\cal Z} = e^{-\beta E_0^0} = e^{-s_2 E_0^0} \times e^{-(s_1-s_2) E_0^0}\times e^{-(\beta-s_1) E_0^0}.
\end{equation}
Here the 1st and 3d factor cancel against the free evolution of the quench problem,
i.e. $e^{-s_2  H_0} |0,k^0_0\rangle = |0,k^0_0\rangle e^{-s_2 E_0^0}$ and the same for
another free 'phase', resulting in 
\begin{equation}
\label{eq:G_f_spectral_low}
G_f(s_{12}) \!=\!\frac{1}{\sqrt{\pi}} \!\!\int\limits_0^{+\infty}\!\! \frac{d\alpha}{\sqrt{\alpha}} 
\sum_{k^\alpha_n}|\langle 0, k^0_0| e^{\phi/4}| \alpha, k^\alpha_n \rangle|^2 
e^{ - (E_n^\alpha-E_0^0) s_{12}}.  \nonumber
\end{equation}
For very large  times,
the sum is dominated by the 0-th level $E_0^\alpha$. Moreover the dominant contribution to the $\alpha$ integral comes from small $\alpha$, where one can use the first order perturbation theory: $E_0^\alpha-E_0^0 \approx \alpha \langle 0, k^0_0| e^{\phi}| 0, k^0_0 \rangle$.    This way one finds at long times $s\gg \Lambda^{-1}$:
\begin{eqnarray}
G_f(s)\! &=& \!\frac{1}{\sqrt{\pi}} \int\limits_0^{+\infty}\! \frac{d\alpha}{\sqrt{\alpha}} 
\langle 0, k^0_0| e^{\phi/4}| 0, k^0_0 \rangle^2 \, e^{ -\alpha \langle 0, k^0_0| e^{\phi}| 0, k^0_0 \rangle s} \nonumber\\
&=&\frac{\langle 0, k^0_0| e^{\phi/4}| 0, k^0_0 \rangle^2}{\langle 0, k^0_0| e^{\phi}| 0, k^0_0 \rangle^{1/2}}\, \frac{1}{s^{1/2}},
\end{eqnarray}
which with the help of Eq.~(\ref{eq:Matrix_El}) yields Eq.~(\ref{eq:fast_G_asym_long_t}).


\end{document}